# Observation of Anti-helical Edge States in Acoustic Metamaterials

Tianzhi Xia,[1,2] Qicheng Zhang,[1] and Chunyin Qiu[1*]

[1]School of Physics and Technology, Wuhan University, Wuhan 430072, China
[2]College of Physics and Electronic Engineering, Qujing Normal University, Qujing 655011, China

[*]To whom correspondence should be addressed: cyqiu@whu.edu.cn

*Abstract.* As a hallmark of the quantum Hall effect, chiral edge modes (CEMs) counter-propagate along the two parallel edges of a ribbon structure. However, recent studies demonstrate counterintuitive anti-CEMs that co-propagate along the parallel edges. Analogous to the established extension of the CEMs to helical edge modes (HEMs) in the quantum spin Hall effect, it is natural to extend the anti-CEMs to anti-HEMs, which comprise a pair of time-reversal-related anti-CEMs. In this Letter, we report the first observation of the anti-HEMs based on a bilayer model that features staggered positive and negative interlayer hoppings. Experimentally, we implement this anti-helical model on an acoustic platform and provide compelling evidence for the anti-HEMs by selectively exciting different spin subspaces, along with identifying the energy-biased Dirac points in bulk spectra. Our findings may offer new insights into topological phases of matter and potentially pave the way for designing novel devices with unique edge transport properties.

*Introduction.* In the past few years, transplanting fundamental concepts from condensed matter physics into artificial crystals has led to substantial progress in topologically protected phenomena [1-3]. Among these, one-way transport—protected by band topology—stands out as the most remarkable feature [4-18]. The early discoveries can be primarily classified into two categories: chiral edge modes (CEMs) and helical edge modes (HEMs). The former are observed in two-dimensional (2D) systems with broken time-reversal (TR) symmetry, such as quantum Hall effects (QHE) and Haldane insulators [4-13]. In these systems, the bulk bands are gapped while gapless edge states propagate in opposite directions along the two parallel edges of a strip structure [Fig. 1(a)]. The latter, HEMs, occur in TR-invariant systems such as the 2D quantum spin Hall effect (QSHE) [4,8,14-18], and can be regarded as two superimposed copies of Haldane insulators related by TR symmetry [Fig. 1(b)]. Both the CEMs and HEMs, being topologically protected and robust against imperfections and disorder, exhibit significant potential for practical applications and are central to the continued growth of interest in topological states of matter [1-3].

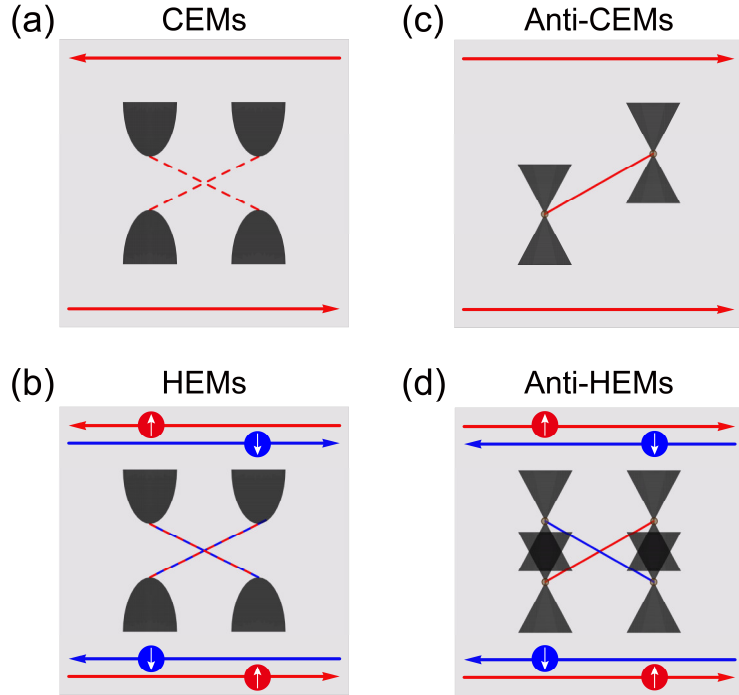

FIG. 1. Schematic illustrations of four distinct types of topological edge modes in 2D systems. (a) Chiral edge modes (CEMs) in a (gapped) QHE system, which counter-propagate along the two parallel edges. (b) Helical edge modes (HEMs) in a (gapped) QSHE system, which consist of two sets of spin-locked CEMs. (c) Anti-CEMs in a modified Haldane model, which co-propagate along the parallel edges. (d) Anti-HEMs

focused in this work, which comprise two sets of spin-locked anti-CEMs.

Recently, a new type of edge modes, dubbed anti-chiral edge modes (anti-CEMs), was proposed based on a modified Haldane model [19]. In sharp contrast to CEMs, the anti-CEMs propagate in the same direction along both parallel edges [Fig. 1(c)], challenging conventional understanding and opening new avenues for exploring topological transport phenomena. Notably, the anti-CEMs only emerge in gapless systems, where the bulk states provide the necessary counter-propagating modes to ensure a legitimate ribbon band structure with full Brillouin zone periodicity. (Note that the number of counter-propagating bulk modes themselves is not balanced within the energy range of the anti-CEMs.) Experimentally, the anti-CEMs have been observed in TR symmetry-broken photonic crystals [20-23] and circuit systems [24], as well as indirectly in TR-symmetric higher dimensions via synthetic gauge fields [25,26]. Inspired by the extension of QHE to QSHE, as illustrated in Fig. 1(d), it is natural to expect spin-enriched anti-HEMs in 2D TR-invariant systems [27]. Compared to anti-CEMs, the anti-HEMs offer enhanced control over wave fields by leveraging the additional spin degree of freedom, thereby opening new possibilities for applications such as robust sensing and communication systems. Despite their great appeal, progress in anti-HEMs has been hindered primarily by the lack of experimentally feasible theoretical models that can guide practical implementation. The key challenge lies in incorporating spin degrees of freedom into the already complex realization of anti-chiral models [20-24], which require breaking TR symmetry—an inherently difficult task in classical systems, especially in passive acoustic systems.

In this Letter, we propose a simple yet universal design strategy for constructing anti-helical models, and report the first experimental observation of anti-HEMs using acoustic metamaterials. More concretely, inspired by the bilayer design of acoustic pseudospins [28-34], we begin with a lattice model consisting of two decoupled, TR-related anti-chiral layers. Each anti-chiral copy, defining a up or down pseudospin, involves purely imaginary intra-layer hopping (see Fig. 2). Through a similarity transformation, we convert the system into a layer-coupled model with only real-valued hoppings. It can be described by a new set of pseudospins, $|\pm\rangle$, as linear combinations of the original spin-up and spin-down states. Importantly, this new model can be realized in passive acoustic metamaterials, where different pseudospin subspaces are distinguishable through selective sound excitations [34]. In our acoustic experiments, we have not only characterized the *spin-dependent* bulk and edge spectra in momentum space, but also demonstrated the co-propagating nature of the anti-HEMs in real space, thereby providing conclusive evidence for the validity of our theoretical model. All experimental results align well with the theoretical predictions. Our work expands the scope of one-way edge transport and opens new avenues for wave manipulations.

*Tight-binding model.* As shown in Fig. 2(a), we start with a simple monolayer model with anti-CEMs. It can be viewed as a deformed graphene lattice featuring a real-valued hopping $\kappa$, combined with an additional purely imaginary hopping $iJ$ (see Fig. S1 in Supplemental Material [35]). Mathematically, the $\boldsymbol{k}$-space Hamiltonian of the system reads

$$H_\uparrow(\boldsymbol{k}) = d_0(\boldsymbol{k})\sigma_0 + d_1(\boldsymbol{k})\sigma_1 + d_2(\boldsymbol{k})\sigma_2, \tag{1}$$

where $d_0(\boldsymbol{k}) = -2J\sin(\boldsymbol{k}\cdot\boldsymbol{a}_1)$, $d_1(\boldsymbol{k}) = \kappa[1 + \cos(\boldsymbol{k}\cdot\boldsymbol{a}_1) + \cos(\boldsymbol{k}\cdot\boldsymbol{a}_2)]$, and $d_2(\boldsymbol{k}) = -\kappa[\sin(\boldsymbol{k}\cdot\boldsymbol{a}_1) + \sin(\boldsymbol{k}\cdot\boldsymbol{a}_2)]$, with $\boldsymbol{a}_1$ and $\boldsymbol{a}_2$ being lattice vectors and $\sigma_i$ being Pauli matrices. Physically, the terms $d_1(\boldsymbol{k})\sigma_1$ and $d_2(\boldsymbol{k})\sigma_2$ are inherited from the graphene lattice, which enforce a pair of stable Dirac points in momentum space. More importantly, the system's chiral symmetry enables quantized winding numbers (for 1D subsystems of constant $\boldsymbol{k}_1$ or $\boldsymbol{k}_2$) and the emergence of flat edge bands that connect the Dirac points for a periodic ribbon structure. The first term, $d_0(\boldsymbol{k})\sigma_0$, is proportional to the unit matrix $\sigma_0$ in the sublattice space. This $\boldsymbol{k}$-dependent pseudo-scalar potential offsets the energies of the Dirac points due to the breaking of TR symmetry (as long as $J \neq 0$). Nevertheless, this term does not alter the spinor structure of the wave functions compared to the pristine graphene lattice [19]. Ultimately, the modified Haldane model exhibits dispersive anti-CEMs, with the band topology encoded in the wave functions. To demonstrate the aforementioned physics, we present the band structure for a system with $\kappa = -1$ and $J = 0.5$ [Fig. 2(b)]. It clearly shows a pair of energy-biased Dirac points along the diagonal line $\Gamma'$-M-$\Gamma''$ in the Brillouin zone: one lies above zero energy, and the other below. When the system is truncated in the $\boldsymbol{a}_2$ direction, energetically-degenerated edge states emerge in the two $\boldsymbol{a}_1$-directed edges and propagate along the same direction, as illustrated by the red lines that connect the two Dirac points.

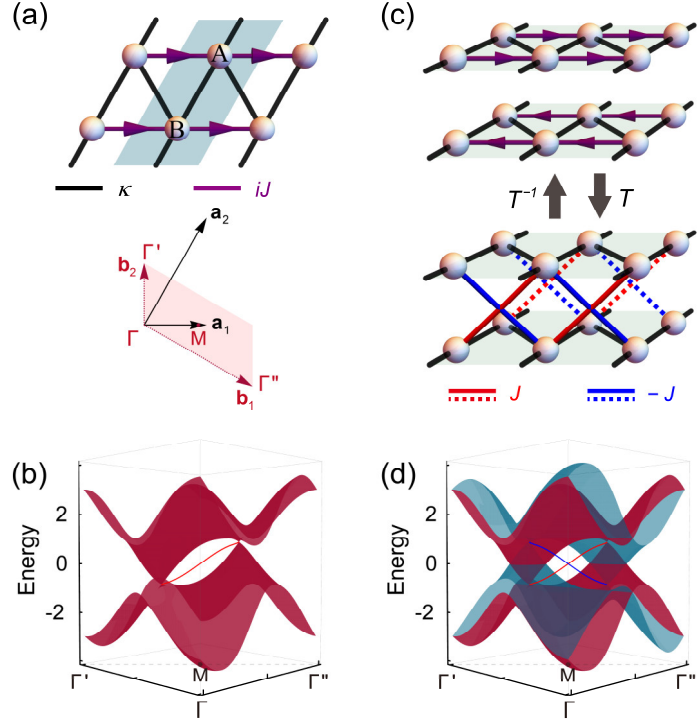

FIG. 2. Tight-binding model. (a) Top panel: Monolayer anti-chiral model. Each unit cell (blue shaded) consists of two inequivalent sites A and B, coupled with the real-valued hopping $\kappa$ (black bonds) and purely imaginary hopping $iJ$ (purple bonds). Bottom panel: Momentum-space Brillouin zone spanned by the reciprocal lattice vectors $\boldsymbol{b}_1$ and $\boldsymbol{b}_2$, with $\boldsymbol{a}_1$ and $\boldsymbol{a}_2$ being their real-space counterparts. (b) Band structure of the anti-chiral model, which features a pair of energy-biased Dirac points along the diagonal of the Brillouin zone, $\Gamma'$-M-$\Gamma''$. The red line linking Dirac points illustrates degenerate anti-CEMs. (c) Schematic diagram of bilayer anti-helical models. Top panel: Anti-helical model consisting of two TR-related, decoupled anti-chiral layers, where the top and bottom ones define pseudospins $|\uparrow\rangle$ and $|\downarrow\rangle$, respectively. Bottom panel: Layer-coupled anti-helical model, obtained via a similarity transformation from the above layer-decoupled one. It features staggered positive (red bonds) and negative (blue bonds) inter-layer couplings, $\pm J$, with solid and dashed lines indicating which sublattices the couplings belong to. Note that new pseudospins after the similarity transformation are labeled as $|+\rangle$ and $|-\rangle$. (d) Band structure of the anti-helical system, where the (colored) bands are classified into two different pseudospin subspaces.

To construct an anti-helical model, intuitively, we consider first a decoupled bilayer model, where the top and bottom layers are characterized by $H_\uparrow(\boldsymbol{k})$ and its TR counterpart $H_\downarrow(\bm{k})$, as illustrated in the top panel of Fig. 2(c). The Hamiltonian of the bilayer system is given by

$$H(\bm{k}) = H_\uparrow(\bm{k}) \oplus H_\downarrow(\bm{k}). \tag{2}$$

As a direct but trivial consequence, the bilayer model hosts a pair of anti-CEMs that propagate in opposite directions on the two decoupled layers. However, this model has limited physical significance, as it essentially consists of two independent anti-chiral layers, not to mention the substantial challenge of realizing purely imaginary hopping in acoustic systems later. To make the model more physically relevant, a similarity transformation $H'(\bm{k}) = TH(\bm{k})T^{-1}$ is performed to couple the two layers and make all the hoppings real-valued, where the transformation matrix is given by $T = \frac{1}{\sqrt{2}}\begin{pmatrix} i & -i \\ 1 & 1 \end{pmatrix} \otimes \sigma_0$. This leads to a coupled bilayer model with Hamiltonian

$$H'(\bm{k}) = -d_0(\bm{k})\sigma_2 \otimes \sigma_0 + d_1(\bm{k})\sigma_0 \otimes \sigma_1 + d_2(\bm{k})\sigma_0 \otimes \sigma_2. \tag{3}$$

More concretely, it corresponds to the lattice model displayed in the bottom panel of Fig. 2(c), where the *purely imaginary* intralayer hoppings $\pm iJ$ in the decoupled bilayer model are transformed into the *real-valued* interlayer hoppings, $\pm J$. (Such positive-negative pairwise couplings have been widely realized in various classical platforms [36-39], including passive acoustic systems [40].) This operation significantly alleviates the challenges encountered in our acoustic experiments. Note that the systems $H(\bm{k})$ and $H'(\bm{k})$ share the same eigenvalues, while the associated eigenvectors are transformed according to the matrix $T$. (In other words, the similarity transformation can be interpreted as observing the same physics from different pseudospin spaces.) Specifically, if we denote the (layer-polarized) pseudospins of $H$ as $|\uparrow\rangle$ and $|\downarrow\rangle$, and the (layer-mixed) pseudospins of $H'$ as $|+\rangle$ and $|-\rangle$, we have the following combination relations: $|+\rangle = \frac{i}{\sqrt{2}}|\uparrow\rangle + \frac{1}{\sqrt{2}}|\downarrow\rangle$ and $|-\rangle = -\frac{i}{\sqrt{2}}|\uparrow\rangle + \frac{1}{\sqrt{2}}|\downarrow\rangle$. Notice that the layer-coupled model $H'$ can be characterized by a projective mirror symmetry, $[\mathcal{M}_z, H'] = 0$, with $\mathcal{M}_z = i\sigma_2 \otimes \sigma_0$ being the projective mirror operator. The pseudospin states $|\pm\rangle$ are distinguishable from the projective mirror eigenvalues $m_z = \pm i$ [34].

Figure 2(d) shows the band structure of the model $H'$, where the red and blue ones correspond to the pseudospins $|+\rangle$ and $|-\rangle$, respectively. When the system is truncated in the $\bm{a}_2$ direction, two pairs of degenerate anti-CEMs (illustrated by blue and red lines) emerge along the $\bm{a}_1$-directed ribbon edges—each co-propagating pair being supported by one of the pseudospin subspaces. Note that the anti-HEMs do not appear along the $\bm{a}_2$-drected ribbon edges (see Fig. S2 in Supplemental Material [35]).

Although exemplified by a simple anti-chiral model [see Eq. (1)], our bilayer construction strategy can also be applied to other monolayer anti-chiral models, such as the well-known modified Haldane model [19].

*Acoustic implementation of the layer-coupled anti-helical model.* The coupled bilayer model $H'$ in Eq. (3) can be readily realized by acoustic metamaterials without breaking TR symmetry, where the orbitals and hoppings are mimicked by air-filled cavity resonators and narrow tubes, respectively [40-43]. As shown in Fig. 3(a), through precise control of the geometry parameters, we have designed a cross-coupled bilayer acoustic structure that corresponds to the tight-binding model. More specifically, the side length of the hexagonal prism cavities is set to $l = 6$ mm and the cavity height is set to $H = 32.9$ mm—the latter results in a dipole mode at 5.25 kHz, well-separated from the other resonance modes. In addition, the connections and lengths of the coupling tubes are carefully designed to achieve the desired sign for both intralayer and interlayer couplings [40]. The cross-sectional area of the intralayer coupling tube is 10.2 mm$^2$, while the cross-sectional areas of the inter-layer coupling tubes are set to 5.1 mm$^2$. Finally, our acoustic structure exhibits effective hoppings $\kappa \approx -212$ Hz and $J \approx 106$ Hz. A close match can be observed between the band structures of the tight-binding model and the acoustic metamaterial (see Fig. S3 in Supplemental Material [35]).

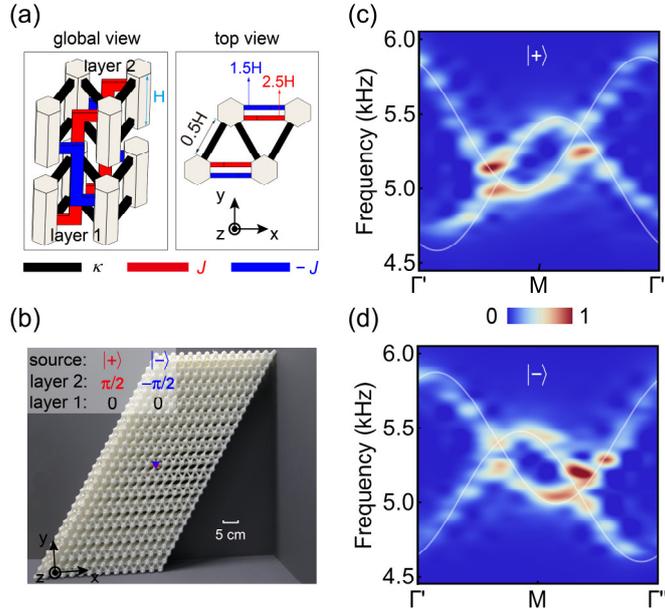

FIG. 3. Acoustic implementation of the layer-coupled anti-helical model and experimental characterization of the energy-biased Dirac points. (a) Acoustic cavity-

tube structure that emulate the tight-binding model. The air-filled cavities mimic the lattice sites, and the connecting tubes of different colors serve as the intra- and interlayer couplings. (b) A photograph of the acoustic sample (top view). To selectively excite the $|\pm\rangle$ pseudospin subspaces, a pair of sound sources, with phase differences of $\pm\pi/2$, are positioned at centers of the top and bottom layers (denoted by colored triangles). (c) Bulk spectrum measured for the $|+\rangle$ pseudospin subspace along the momentum path $\Gamma'$-M-$\Gamma''$. The experimental data (color), showing a pair of energy-biased Dirac points, agree excellently with the tight-binding predictions (white lines). (d) Similar to (c), but for the $|-\rangle$ pseudospin, where the energy shift of the Dirac points is opposite to that of the $|+\rangle$ subspace.

Figure 3(b) shows our experimental setup. The sample consists of $15 \times 15$ unit cells in total, which was 3D-printed using a photosensitive resin material at a fabrication error of ~0.1 mm. To excite and detect the sound waves inside the cavities, small holes were perforated in the cavity resonators for inserting the sound source or probe, which were sealed when not in use. Furthermore, to experimentally resolve the spin-locked bulk (or edge) modes, a pair of identical broadband point-like sound sources was inserted into the bulk (or edge) of the sample, with one positioned in the top layer and the other in the bottom layer. In particular, according to the combination relations $|\pm\rangle = \pm\frac{i}{\sqrt{2}}|\uparrow\rangle + \frac{1}{\sqrt{2}}|\downarrow\rangle$, the top-layer source is applied with a phase delay of $\pm\pi/2$ relative to the bottom-layer one, enabling selective excitation of the pseudospin subspaces $|\pm\rangle$.

Before identifying the highly intriguing anti-HEMs, we first experimentally characterized the bulk properties of our acoustic metamaterial. To do this, we placed a pair of selective acoustic sources at the center of the sample. In our acoustic experiments, we scanned the sound pressure fields cavity by cavity, accompanied by another identical sound probe placed outside the sample for phase reference. Both the input and output signals were recorded and frequency-resolved with a multi-analyzer system (B&K Type 3560B). Through a 2D spatial Fourier transform, we obtained the spin-resolved bulk spectra in momentum space. The experimental data (color scale) in Figs. 3(c) and 3(d), normalized to their respective maximum values, capture well the theoretical band structures (white lines). In particular, the bulk spectrum of the $|+\rangle$ pseudospin subspace exhibits a pair of energy-biased Dirac points along the diagonal $\Gamma'$-M-$\Gamma''$ in the 2D Brillouin zone. A similar phenomenon appears in the bulk spectrum of the $|-\rangle$ pseudospin subspace, but with an opposite energy bias compared to that in the $|+\rangle$ pseudospin subspace. The *spin-dependent* energy bias of the Dirac points provides direct *bulk* evidence for our anti-HEM acoustic metamaterial.

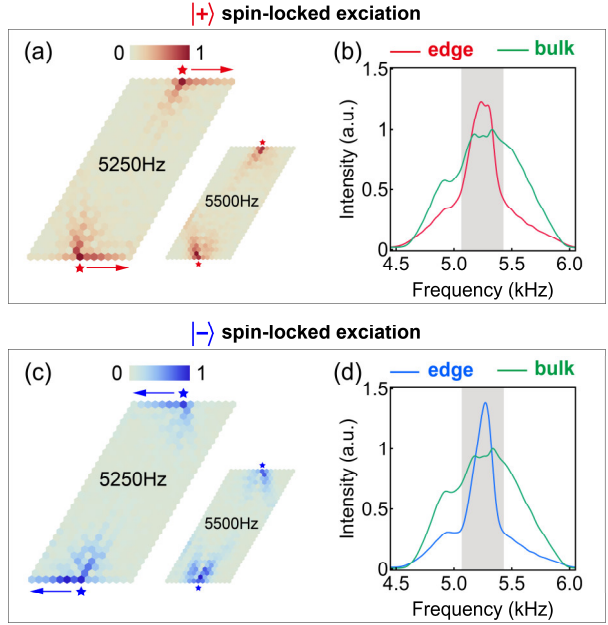

FIG. 4. Real-space observation of anti-HEMs. (a) 2D real-space sound patterns scanned for two typical frequencies. The data are excited by two pairs of $|+\rangle$ spin-locked sources, one positioned at the upper edge while the other at the lower edge. (b) Corresponding sound energy spectra counted separately for bulk and edge sites. The shaded region highlights the frequency window between the biased Dirac points. (c) and (d): Similar to (a) and (b), but excited by two pairs of $|-\rangle$ spin-locked sound sources.

Now we turn to the experimental observation of anti-HEMs. First, we placed two pairs of $|+\rangle$ spin-locked sound sources simultaneously at the middle of the upper and lower sample edges, and scanned the sound profile across the sample. Figure 4(a) exemplifies the sound patterns measured for two typical frequencies. At 5500 Hz, the data show clear sound emission into the bulk, whereas at 5250 Hz, the sound field displays highly localized edge states that propagate rightward along both the *x*-directed edges—this is a hallmark of the $|+\rangle$ spin-locked anti-CEMs. Notably, the sound field at 5250 Hz also shows visible excitations of bulk modes. The coexistence of edge and bulk modes within the frequency range between the biased Dirac points (shaded in gray) can be seen more clearly in Fig. 4(b), which demonstrates the sound energy spectra collected separately for the edge and bulk sites. Similar phenomena can be observed in Figs. 4(c) and 4(d), which provide the data excited by two pairs of $|-\rangle$ spin-locked sound sources. In this case, the anti-CEMs co-propagate along the −*x*-direction as expected. These results, aligning well with the full-wave simulations and Green's function calculations (see Figs. S4 and S5 in Supplemental Material [35]), together visualize the presence of anti-HEMs that consist of a pair of TR-related anti-CEMs.

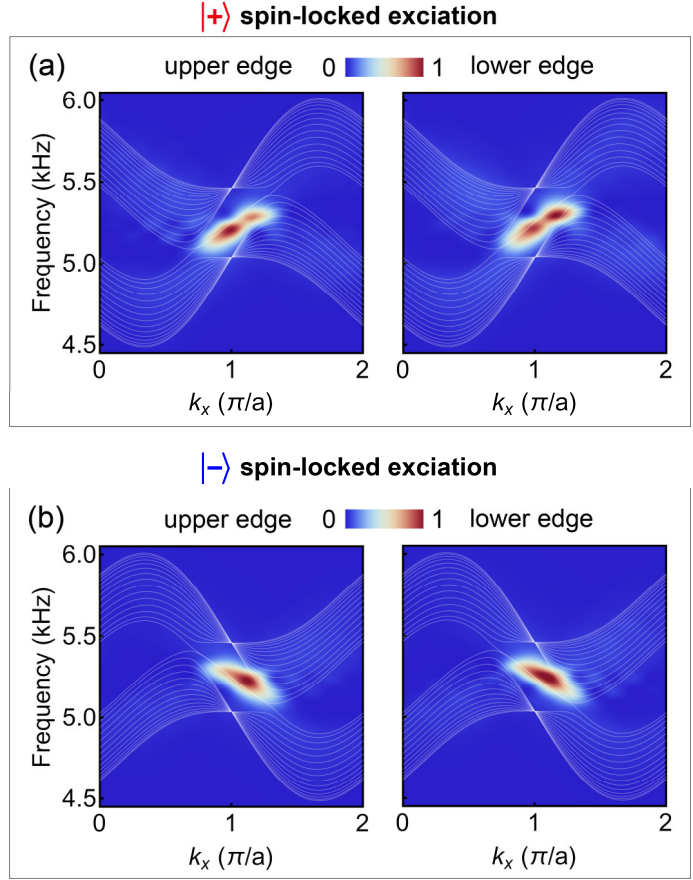

FIG. 5. Experimental spectral evidence for the existence of anti-HEMs. (a) Edge spectra (color) implemented for the upper (left panel) and lower (right panel) edges, both excited by $|+\rangle$ spin-locked sound sources. The white lines represent projected edge spectra calculated using the tight-binding model. (b) Similar to (a), but excited by $|-\rangle$ spin-locked sound sources. As expected, the excited anti-CEMs in the $|+\rangle$ pseudospin subspace demonstrate positive slopes for both the upper and lower edges, while those in the $|-\rangle$ pseudospin subspace exhibit negative slopes.

To further identify the anti-HEMs composed of spin-locked anti-CEMs, we performed 1D spatial Fourier transforms to the sound pressure fields extracted along the upper and lower edges. Figure 5(a) presents the data excited by the $|+\rangle$ spin-locked sound source. As expected, both the selectively-excited edge spectra exhibit positive slopes within the frequency ranges of edge modes, a key signature of the right-moving anti-CEMs in the $|+\rangle$ spin-locked subspace. For comparison, Fig. 5(b) shows the case of $|-\rangle$ spin-locked excitation, which exhibits the left-propagating anti-CEMs with negative slopes for both edges. Combined with the real-space sound patterns in Figs. 4(a) and 4(c), the edge spectra provide conclusive experimental evidence for the presence of anti-HEMs in our gapless acoustic metamaterial.

*Conclusion.* We introduce and experimentally demonstrate the anti-HEMs in an acoustic metamaterial platform. Intriguingly, our experiments reveal that the spin-locked edge modes co-propagate along the parallel edges of the ribbon sample. This unique edge transport enables spin purification by distinguishing anti-HEMs of different pseudospins—permitting only the desired spin to unidirectionally propagate along both edges while filtering out the other. Note that the anti-HEMs exhibit weaker protection compared to conventional topological edge modes in gapped systems: like all topological semimetals, they can scatter into the bulk in the presence of edge defects or disorder (see Supplemental Material [35], Figs. S6-S7). Although the demonstration focuses on acoustic systems, our proposed anti-helical model—which involves only real-valued couplings—is applicable to a wide range of other classical platforms, including microwave systems [36], photonic systems [37], mechanical systems [38], and topolectrical circuits [39]. These findings pave the way for the development of novel devices that utilize spin-purified unidirectional transport and advanced wave manipulation techniques across various physical platforms.

**Acknowledgements**. This project was supported by the National Natural Science Foundation of China (Grants No. 12374418), the National Key R&D Program of China (Grant No. 2023YFA1406900), and the Fundamental Research Funds for the Central Universities.